\documentclass[english,showpacs,floatfix,11pt]{revtex4}
\usepackage[dvips]{graphicx}
\usepackage{longtable}
\usepackage{amsmath,amssymb}
\usepackage{dcolumn}
\usepackage{latexsym}
\usepackage{babel}
\usepackage[latin1]{inputenc}
\usepackage{color}
\usepackage[colorlinks]{hyperref}
\def\pa{\partial}
\def\al{\alpha}
\def\ga{\gamma}
\def\dl{\delta}

\def\kp{\kappa}
\def\th{\theta}
\def\sg{\sigma}
\def\vf{\varphi}

\def\eps{\varepsilon}
\def\l{\left}
\def\r{\right}
\def\nn{\nonumber}

\begin{document}
\title{A Procedure to Solve the Eigen Solution to Dirac Equation}
\author{Ying-Qiu Gu}
\email{yqgu@fudan.edu.cn} \affiliation{School of Mathematical
Science, Fudan University, Shanghai 200433, China} 
\date{23rd August 2017}

\begin{abstract}
In this paper, we provide a procedure to solve the eigen solutions
of Dirac equation with complicated potential approximately. At
first, we solve the eigen solutions of a linear Dirac equation with
complete eigen system, which approximately equals to the original
equation. Take the eigen functions as base of Hilbert space, and
expand the spinor on the bases, we convert the original problem into
solution of extremum of an algebraic function on the unit sphere of
the coefficients. Then the problem can be easily solved. This is a
standard finite element method with strict theory for convergence
and effectiveness.

\end{abstract}
\maketitle

\section{Introduction}
\setcounter{equation}{0}

To study the properties of elementary particles we solve the eigen
solutions of Dirac equations. In some cases with symmetrical
potential, the eigen solutions of Dirac can be solved
exactly\cite{1}- \cite{gu}. However, in the usual cases, the
rigorous solution is absent, and we have to solve the approximate
solutions with required accuracy. Quantum field theory provides a
method to solve the approximate solutions. However it is
inconvenient for some cases due to the complicated procedure,
infinity problem.

In this paper, we provide a standard finite element method to solve
the eigen solutions approximately, which is efficient for most cases
and can be easily realized by computer. The solving procedure is
that, at first, we solve the eigen solutions ${\psi_n}$ of a linear
Dirac equation with complete eigen functions, which approximately
equals to the original equation. The normalized eigensolutions
${\psi_n}$ form the bases of Hilbert space, and we can represent the
solutions of the original Dirac equation by $\phi=\sum X_n \psi_n ,$
where $X_n$ are coefficients. Substituting it into the action of the
original equation, we convert the problem into solving the extremum
of an algebraic equation on the unit sphere $\sum X_n^2=1$, which is
much simpler than the original one. The calculation shows that this
procedure is effective and convenient and suitable for nonlinear
Dirac equations. This process is similar to second quantization, but
here only normal mathematics is involved and the calculation can be
easily performed by computer. In what follows, we take an electron
in Coulomb potential and magnetic field as an example to show the
solving procedure.

\section{Equations and  Simplification}
\setcounter{equation}{0}

At first, we introduce some notations. Denote the Minkowski metric
by $\eta_{\mu\nu}={\rm diag}(1,-1,-1,-1)$, Pauli matrices by
\begin{eqnarray}
 {\vec\sg}=(\sg^{j})= \l \{\l (\begin{array}{cc}
 0 & 1 \\ 1 & 0 \end{array} \r),\l (\begin{array}{cc}
 0 & -i \\ i & 0 \end{array} \r),\l (\begin{array}{cc}
 1 & 0 \\ 0 & -1 \end{array} \r)
 \r\}.\label{1.1}\end{eqnarray}
Define $4\times4$ Hermitian matrices as follows
\begin{eqnarray}\al^\mu=\l\{\l ( \begin{array}{cc} I & 0 \\
0 & I \end{array} \r),\l (\begin{array}{ll} 0 & \vec\sg \\
\vec\sg & 0 \end{array}
\r)\r\},\qquad \ga =\l ( \begin{array}{cc} I & 0 \\
0 & -I \end{array} \r), \label{1.2}
\end{eqnarray}
where $\mu\in\{0,1,2,3\}$, $x^0=ct$. The Dirac equation for an
electron in potential $A^\mu$ is given by
\begin{eqnarray}
\al^\mu(\hbar i\pa_\mu-e A_\mu)\phi &=&\mu c\ga\phi, \label{drc}
\end{eqnarray}
in which the potential reads
\begin{eqnarray}
A_0=-\frac {Ze} r,~~\vec A=\frac 1 2 B(-y,x,0)=\frac 1 2 Br\sin
\th(-\sin\vf, \cos\vf,0).
  \label{ptn} \end{eqnarray}
The corresponding Lagrangian is given by
\begin{eqnarray}
{\cal L}&=&\phi^+\al^\mu (i\pa_\mu-e A_{\mu})\phi-\mu c \phi^+
\ga\phi. \label{lag}
\end{eqnarray}
If the magnetic field $B\ne 0$, the rigorous solution is absent.

In this case, the magnetic quantum number $m_z$ and the sipn $s$ are
still conserved, which also hold for most cases. So the eigen
solution takes the following form\cite{gu}
\begin{eqnarray}
\phi=(u_1,u_2e^{\vf i},-iv_1,-iv_2e^{\vf i})^T \exp(m_z \vf
i-\frac{mc^2} \hbar it),  \label{fms}
\end{eqnarray}
where the index `T' stands for transpose, $ m_z \in\{0,\pm 1,\pm
2,\cdots\}$, and $u_k,v_k(k=1,2)$ are real functions of $r$ and
$\th$.

In the case $B=0$, we have $u_1=u_2,v_1=\pm v_2$ and the solution
can be solved in the form of spin spherical harmonics\cite{1}.
However this solution has complicated coefficients, which is
inconvenient for expansion as bases of Hilbert space.

In order to simplify (\ref{lag}) for approximate computation, we
make transformation
\begin{eqnarray} g=u_1+u_2 i\qquad f=v_1-v_2 i.
\label{gf}
\end{eqnarray}
Substituting it into (\ref{lag}) we get
\begin{eqnarray}
{\cal L}  =  ({\cal L}_0 +{\cal L}_{B} +{\cal L}_f)\mu c,
\label{lagt}
\end{eqnarray}
\begin{eqnarray}
\begin{array}{lll}
 {\cal L}_0 &\equiv & \l(\Re \left[e^{\th i}\left( -\bar g
(\pa_r+\frac i r \pa_\th) f + f (\pa_r+\frac i r \pa_\th) \bar g
\right)\right]
-\frac i {r\sin\th}(m_z+\frac 1 2 ) (\bar g\bar f-gf)\r)\rho  \\
&&+\l( \eps (|g|^2+|f|^2) -\frac {Z\al\rho}r |g|^2 -(2+\kp)|f|^2
\r), \end{array} \label{lag0}
\end{eqnarray}
\begin{eqnarray}
{\cal L}_f &\equiv & \l(\kp-\frac {Z\al\rho}r\r)|f|^2, \label{lage}\\
 {\cal L}_B &\equiv & \mu_B B\cdot
\frac{ i (gf-\bar g \bar f)r\sin\th }{\rho },
\label{lagb}\end{eqnarray} in which $\eps\ll 1$ is dimensionless
energy defined by $m=(1-\eps)\mu$, $\rho=\frac \hbar {\mu c}$ is the
Compton wave length used as length unit, $\kp$ is a constant to
improve convergent rate. In the case (\ref{ptn}) we set $\kp=0$ due
to the small value of $\alpha$ or weakness of electromagnetic
interaction. For the strong interaction we can set $\kp$ equal to
the average potential\cite{prb}. $\mu_B =\frac {\hbar e} {2m}$ is
the Bohr magneton of electron.

In (\ref{lagt}), ${\cal L}_0$ almost keeps all invariance of
relativity and has simple and complete eigensolutions, which can be
used as the bases of Hilbert space, we call it the representation
space of spinor. ${\cal L}_f$ and ${\cal L}_B$ are the trouble terms
with small energy, which act as perturbation in the calculation.

In what follows we take $\mu c=1$ as energy unit, then (\ref{lagt})
becomes dimensionless.  For (\ref{lag0}), we can solve the rigorous
eigensolutions by making transformation
\begin{eqnarray}
g=U(r) M(\th),~~f=V(r) N(\th),~~M=P(\th)+Q(\th)i.
\label{gfuv}\end{eqnarray} By variation of (\ref{lag0})  we find
$N=M e^{-i\th}$ and
\begin{eqnarray}
\pa_\th P &=& \cot\th m_z P+(m_z+K)Q,\label{eqp}\\
\pa_\th Q &=& -\cot\th(m_z+1)Q+(m_z+1-K)P ,
\label{eqq}\end{eqnarray} in which $K=\pm1,\pm2,\cdots$
corresponding to orbital angular momentum, $P,Q$ are associated
Legendre functions. The radial functions satisfy
\begin{eqnarray}
\pa^2_r U +\frac 2 r \pa_r U -\l(\frac {K(K-1)}{r^2}+\frac
{\eps(2-\eps)}{\rho^2}+\frac{Z\al(2-\eps)}{r\rho}\r)U=0 ,
\label{equ}\end{eqnarray} and
\begin{eqnarray}
V=\frac{\l(r\pa_r U-(K-1)U+\r)\rho}{(2-\eps)r}.
\label{eqv}\end{eqnarray} The above equations can be easily solved,
and the solutions are all elementary functions. The normalizing
conditions are as follows
\begin{eqnarray}
\int_0^\pi (P^2+Q^2) 2\pi \sin\th d\th=1,\quad \int_0^\infty
(U^2+V^2) r^2 d r=1.  \label{eqv}\end{eqnarray}

\section{Eigen Solutions to the equation}
\setcounter{equation}{0}

Due to the parity invariance of the eigensolutions, if $g$ takes the
form $\sum U_n \exp(2n\th i)$, then $f$ will be $\sum V_n
\exp((2n-1)\th i)$, or vice verse. Considering the case $m_z=0$, by
solving (\ref{eqp}) and (\ref{eqq}), we have normalized functions
$M_{K}$. Some simple ones are displayed as follows
\begin{eqnarray}
M_{-2}=\frac {\sqrt 2} {8\sqrt\pi}(3e^{2\th i}+1),~~M_1=\frac 1
{2\sqrt{\pi}},~~M_3 =\frac {\sqrt{3}} {16 \sqrt{\pi}} (e^{2\th
i}+2+5 e^{-2\th i}). \label{slm}\end{eqnarray} In usual cases,
taking $|K|\le 3$ is enough for approximate solution. By
(\ref{equ}), we get
\begin{eqnarray}
\begin{array}{lll}
U&=&\l[ C_1 r^{K-1} L_{n-K}^{2K-1}\l(\frac {2r} {r_n}\r) + C_2
r^{-K} L_{n+K-1}^{1-2K}\l(\frac {2r} {r_n}\r) \r] \exp\l({-\frac r
{r_n}}\r),
\end{array}  \label{slu}\end{eqnarray}
where $L$ is Laguerre polynomials, $n\ge |K|$ is positive integer,
$C_1=0$ corresponding to $K<0$ and $C_2=0$ corresponding to $K>0$,
and
\begin{eqnarray}
\eps_n = \frac {2Z^2 \al^2} {Z^2\al^2+4n^2}, \quad r_n =\frac
{(Z^2\al^2+4n^2)\rho} {4Z\al n}. \label{eps}\end{eqnarray}
Substituting (\ref{slu}) and (\ref{eps}) into (\ref{eqv}), we can
get function $V$. For all eigensolutions we have
\begin{eqnarray}
\int_0^\infty U^2_{K,n} r^2 d r=1-\frac 1 2 \eps_n,\quad
\int_0^\infty V^2_{K,n} r^2 d r=\frac 1 2 \eps_n.
\label{uvn}\end{eqnarray} Since the solution $U_{K,n}$ can be easily
generated by computer, here we only display the simplest one
\begin{eqnarray}
U_{1,1}=\frac {\sqrt{2(2-\eps_1)}}{\sqrt{r_1^{3}}} e^{ \frac {r}
{r_1}},\quad V_{1,1}=\frac{-\sqrt 2 \rho}{\sqrt{(2-\eps_1)r_1^5}}
e^{-\frac r {r_1}}. \label{uv11}\end{eqnarray}

Due to the normalization of the eigenfunctions, the calculation of
expansion of $(g,f)$ is convenient. For example, we take
\begin{eqnarray}
g &=& \sum_{k=0}^3 X_k U_{1,k+1} M_1 +(X_4 U_{-2,3}+X_5
U_{-2,4})M_{-2}+(X_6 U_{3,3}+X_7 U_{3,4})M_{3}.
\label{gfg}\\
f &=& [\sum_{k=0}^3 X_k V_{1,k+1} M_1 +(X_4 V_{-2,3}+X_5
V_{-2,4})M_{-2}+(X_6 V_{3,3}+X_7 V_{3,4})M_{3}] e^{-\th i}.
\label{gff} \end{eqnarray} Substituting them into (\ref{lag0}) we
have the action
\begin{eqnarray}
I_0 \equiv 2\pi \int_0^\infty dr\int_0^\pi{\cal L}_0 r^2\sin\th
d\th=\sum_{k=0}^7(\eps-\eps_{k})X_k^2, \label{i0}
\end{eqnarray}
which is diagonal due to eigenfunctions of ${\cal L}_0$. This can be
used to check the correctness of computing program. Usually, $(g,f)$
is mainly related with the eigenfunctions whose quantum numbers near
that of $(g,f)$.

Substituting (\ref{gfg}, \ref{gff}) into (\ref{lage}), we can get
action $I_e=-\sum a_{k,l} X_kX_l$, and then we have
\begin{eqnarray}
I_0 +I_e =\sum_{k=0}^7(\eps-\eps_{k})X_k^2-\sum_{k,l=0}^7 a_{k,l}
X_kX_l, \label{i0e}
\end{eqnarray}
which becomes the approximate action of an electron in Coulomb
potential. Solving the eigenvalues of (\ref{i0e}) we get the
numerical energy spectrums. Comparing them with rigorous one
\begin{eqnarray}
\eps_{K,n} &=& 1-\l(1+\frac
{\al^2}{\l(n-|K|+\sqrt{K^2-\al^2}\r)^2}\r)^{-\frac 1 2} \nn\\
&=&_{}\frac {\al^2}{2 n^2} +\l(\frac 1 {2n^3|K|} -\frac 3 {8
n^4}\r)\al^4+{\rm O}\l(\al^6\r), \label{rge}
\end{eqnarray}
we have the  accuracy ${\rm O}(\al^6)$ for (\ref{gfg}, \ref{gff}).

Now we compute the magnetic energy of (\ref{lagb}). Substituting
(\ref{slu}, \ref{eps}, \ref{uvn}) and (\ref{eqv}) into ${\cal L}_B$,
we get the energy of magneton
\begin{eqnarray}
\dl E=g_s \mu_B B,\qquad g_s=-\frac {K(2m_z+1)}{2K-1}, \label{gs}
\end{eqnarray}
$g_s$ is similar to the Lande factor, which is independent of $n$.
From the above equations we find that, for the eigenfunctions of
${\cal L}_0$ the relations become simple and neat. So these
eigenfunctions form a good coordinate system for expansion of the
original functions $(g,f)$.

Substituting (\ref{gfg}) and (\ref{gff}) into ${\cal L}_B$ we get
action
\begin{eqnarray}
I_B=-\mu_B B \sum_{k,l=0}^7 b_{k,l}(Z) X_kX_l. \label{ib}
\end{eqnarray}
We solve the eigen values of the coefficient matrix, and then we can
compute the anomalous magnetic moment of a free electron. In this
case, the eigenfunctions just act as the bases of representation
space, rather than the electron is really in Coulomb potential. By
adjusting parameter $Z$, when $Z\dot = 12$ and $\bar r \dot =45\rho$
we get the magnetic moment $g_s=-1.001159652$, which means the wave
function of a free electron is  a concentrated package. However the
magnetic moment of an electron is not a constant, which depends on
its state.

The total approximation action corresponding to the original
equation (\ref{drc}) is given by
\begin{eqnarray}
I=I_0+I_e+I_B=\sum_{k=0}^7(\eps-\eps_{k})X_k^2-\sum_{k,l=0}^7
(a_{k,l}+\mu_B B b_{k,l}) X_k X_l. \label{it}
\end{eqnarray}
Solving the eigenvalues and eigenvectors of the coefficient matrix,
and substituting them into approximation (\ref{gfg}, \ref{gff}), we
get the approximation solutions to the original problem (\ref{drc}).
This process is equivalent to solving the extremum of (\ref{it}) on
the sphere $\sum X_k^2=1$, which is also suitable for the case with
nonlinear potentials.

\section{Discussion and Conclusion}
\setcounter{equation}{0}

In this paper we provide a convenient procedure to approximately
solve the eigen solutions to the Dirac equation with complicated
potentials. (\ref{lag0}) has complete eigen functions and
(\ref{equ}) is similar to the Schr\"odinger equation. The
approximate equation (\ref{lag0}) keeps all main properties of the
original equation (\ref{lagt}), such as energy spectrums, invariance
etc. Expressing the physical variables and relations of spinor by
the eigen functions of the representation space, we have simple and
neat formalism, such as (\ref{slu}, \ref{eps}, \ref{uvn}) and
(\ref{gs}).

This procedure is a standard finite element method which has strict
mathematical theory for its convergence and effectiveness. Practical
simulation shows the procedure is also suitable for computing
nonlinear potentials. The procedure can be easily realized by
computer. For the total action of the original problem similar to
(\ref{it}), we can design high convergent speed numerical program.

It should be mentioned the base functions as an ensemble have
orthogonality for different $(n,K)$, but the radial base functions
corresponding to different $K$ are not definitely orthogonal.
Besides, under what conditions the bases of the representation space
have completeness is still a problem.


\begin{thebibliography}{99}
\bibitem{1} W. Greiner, {\em Relativistic Quantum Mechanics (Wave Equations)}, Springer-Verlag Berlin Heidelberg,
1990
\bibitem{2} G. V. Shishkin, V. M. Villalba,{\em  Electrically neutral Dirac
particles in the presence of external fields: exact solutions}, J.
Math. Phys. 34 (1993) 5037-5049, hep-th/9307061
\bibitem{3} V. G. Bagrov, M. C. Baldiotti, D. M. Gitman, I. V. Shirokov,
{\em New solutions of relativistic wave equations in magnetic fields
and longitudinal fields}, J. Math. Phys. 43 (2002) 2284-2305,
hep-th/0110037
\bibitem{5} M. A. Rodriguez, P. Winternitz,{\em  Quantum Superintegrability
and Exact Solvability in N Dimensions}, J. Math. Phys. 43 (2002)
1309-1322, math-ph/0110018
\bibitem{6} P. Winternitz, I. Yurdusen,{\em  Integrable and superintegrable
systems with spin}, J. Math. Phys. 47, 103509 (2006),
math-ph/0604050
\bibitem{gu} Y. Q. Gu, {\em Integrable conditions  for  Dirac Equation and Schr\"odinger
equation}, arXiv:0802.1958
\bibitem{prb} Y. Q. Gu, {\em Mass Spectrum of Dirac Equation with Local Parabolic Potential}, arXiv:hep-th/0612214
\end{thebibliography}
\end{document}